\documentstyle[eqsecnum,aps,prb]{revtex}

\begin{document}
\draft
\title{Reply to Comment on: ''Exact solutions of the Lawrence-Doniach model for
layered superconductors''}
\author{Sergey V. Kuplevakhsky}
\address{Department of Physics, Kharkov National University,\\
61077 Kharkov, Ukraine\\
and Institute for Low Temperature Physics and Engineering,\\
61103 Kharkov,\\
UKRAINE}
\date{\today}
\maketitle

\begin{abstract}
In the recent Comment [V. M. Krasnov, Phys. Rev. B {\bf 65}, 096503 (2002)],
the author claims to have ''disproved'' our theoretical conclusion [S. V.
Kuplevakhsky, Phys. Rev. B {\bf 60}, 7496 (1999); {\it ibid.} B {\bf 63},
054508 (2001)] that isolated Josephson vortices in layered superconductors
and stacked junctions are absolutely unstable in the presence of an external
field. We show that this claim has no grounds. Moreover, by solving an
appropriate boundary value problem, we obtain a complete classification of
soliton (vortex) solutions to coupled static sine-Gordon equations. We also
discuss the problem of vortex penetration and analyze available experimental
data.
\end{abstract}


Unfortunately, the criticism contained in the recent Comment by Krasnov\cite
{Kr02} is mostly a result of a misunderstanding of major physical ideas and
mathematical methods of my papers.\cite{K99,K01} And what is more, ''the
three main arguments against fluxons'', ascribed to me in the Comment,
cannot be found in Refs. \cite{K99,K01}: (i) my proof of the absence of
single-vortex solutions does not appeal to the ''argument of lower free
energy''; (ii) my remark in Ref. \cite{K01} on the problem of the existence
and uniqueness concerns the difference between a finite and an infinite sets
of coupled static sine-Gordon (SG);\cite{r3} (iii) according to Refs. \cite
{K99,K01} , isolated Josephson vortices do not penetrate layered
superconductors in a static, homogeneous external field $H>0$ because they
cannot exist inside, not vice versa. In view of this obvious
misrepresentation of the approach of Refs. \cite{K99,K01} , I have to remind
the key results of these papers and the methods of their derivation.

The actual main arguments of my papers are exact minimization of the Gibbs
free-energy functional of the infinite (in the layering direction) layered
superconductor in the framework of both the Lawrence-Doniach (LD) model
(Ref. \cite{K01}) and the true microscopic model.\cite{K99} In particular, I
resolve a nontrivial mathematical problem of the minimization with respect
to the phases of the superconducting (S) layers. The necessity of this
procedure was realized earlier too.\cite{BC91,BCG91,BLK92,Ko93} However, the
fact that the phases and the vector potential are subject to an intrinsic
constraint relation\cite{K99,K01} (the current-conservation law) was not
noticed in Refs. \cite{BC91,BCG91,BLK92,Ko93} , which led to a failure. The
clarification of this major issue in my papers required the use of rather
sophisticated and rigorous methods of mathematical physics. As is emphasized
in Refs. \cite{K99,K01} , the general equations derived therein encompass
the whole physics of layered superconductors in parallel magnetic fields in
the range $0\leq H\leq H_{c2}$, including all well-known, experimentally
verified limiting cases.

In contrast to the authors of Refs. \cite{B73,CC90,BC91,BCG91,BLK92,Ko93}
(and the author of the Comment too), I do not make any a priori assumptions
about the vortex structure: Vortex-plane solutions along with the Meissner
solution appear as a unique result of exact minimization of the free-energy
functional, which automatically proves their stability and the absence of
any other vortex solutions at $H>0$. Recent experiments\cite{N96,Yu98} on
artificial stacked junctions clearly confirm the existence and stability of
coherent Josephson-vortex configurations that can be unambiguously
identified with vortex planes at $H>0$. Exceptional stability of vortex
planes in the dynamic regime is confirmed by numerical simulations in Refs. 
\cite{SBP93,Kl94} (the authors of these papers use the terms ''in-phase''
and ''phase-locked'' solutions).

Unfortunately, the method of exact minimization of the free-energy
functional, employed in Refs. \cite{K99,K01} , was not understood by the
author of the Comment. We therefore present an alternative, simpler method
in Appendix A.

The claims of the author of the Comment that he has ''disproved'' our
''statements'' have no grounds for one more serious reason. In contrast to
Refs. \cite{K99,K01} , the author of the Comment concentrates on the
discussion of finite Josephson-junction stacks and commits a typical
mistake: Based on his own numerical calculations of the shape of the fluxon
for $H=0$,\cite{Kr01} he makes unjustified extrapolation of the results of
Ref. \cite{Kr01} to the case $H>0$. The existence of single-vortex
configurations at $H>0$ is assumed as ''obvious'', without any mathematical
proof (and even without an exact mathematical definition). However, the
problem of the classification of vortex configurations in finite Josephson
systems belongs to the theory of coupled SG equations and cannot be solved
by means of simplistic ''energy arguments'', numerical simulations or
combinatorics.\cite{Kr02}

In section II, we show that our conclusions\cite{K99,K01} hold for finite
Josephson-junction stacks as well: Single-vortex configurations do not exist
at $H>0$ in this case, either. However, a rigorous proof of this fact is
more involved than in the case of infinite layered superconductors\cite
{K99,K01} and requires different mathematical techniques.\cite{K0,K2} (For
this reason, finite Josephson-junction stacks were not considered in Refs. 
\cite{K99,K01} .) Given that analytical properties of coupled SG equations
have not been studied in any previous publications, we provide a necessary
mathematical background\cite{K0,K2} in section I. In sections III, IV, we
prove the inconsistency of the rest of the arguments of the Comment. In
parallel, we clarify misleading statements of the author of the Comment on
certain important physical and mathematical issues.

For definitiveness, we will adhere to the notation of Ref. \cite{K01} with a
minor exception: the phase difference between the S-layers $n$ and $n-1$
will be denoted as $\phi _n\equiv \varphi _n-\varphi _{n-1}$, with $\varphi
_n$ being the phase. The geometry of the problem is that of figures 1 and 2
in Ref. \cite{K01}: The layering axis is $x$, with $p$ being the period and $%
N$ being the number of S-layers; the $y$ axis is directed along the
S-layers, with $-L\leq y\leq +L$ being the region occupied by the system [or 
$-\infty <y<+\infty $, if $L=\infty $]. The static, homogeneous external
field is applied along the $z$ axis: ${\bf H}=(0,0,H\geq 0)$. The external
current is not considered, i.e., $I=0$.

\section{Some background information on analytical properties of coupled
static SG equations}

The difference in mathematical description of the finite and the infinite LD
models, emphasized in Ref. \cite{K01} , is not a ''speculation'':\cite{Kr02}
It results from translational invariance of the barrier potential and the
absence of external boundaries in the layering direction in the latter case.
The additional symmetry of the infinite model manifests itself in the
appearance of an unphysical degree of freedom. The elimination of this
unphysical degree of freedom requires minimization of the Gibbs free energy
with respect to the phases $\varphi _n$. In contrast, minimization with
respect to $\varphi _n$ is prohibited for the finite model by the physical
requirement that the local field be equal to $H$ at $x=0$, $x=\left(
N-1\right) p$.

As an illustration, we compare coupled static SG equations for $\phi _n$ in
both the cases. [For $r(T)\ll 1$ and $H\ll H_{c2}$, these equations
constitute solvability conditions for the Maxwell equations.\cite{K99,K01}]
For $N=\infty $, they read 
\begin{equation}  \label{1}
\lambda _J^2\frac{d^2\phi _n}{dy^2}=\frac 1{\epsilon ^2}\left[ \left(
2+\epsilon ^2\right) \sin \phi _n-\sin \phi _{n+1}-\sin \phi _{n-1}\right] ,
\end{equation}
\begin{equation}  \label{2}
n=0,\pm 1,\pm 2,\ldots
\end{equation}
Physical solutions to (\ref{1}), (\ref{2}) are subject to the requirement
that the local field be equal to $H$ at $y=\pm L$, which yields\cite{K99,K01}
\begin{equation}  \label{3}
\frac{d\phi _n}{dy}\left( \pm L\right) =2epH.
\end{equation}
Conditions (\ref{3}) allow only for solutions of the type 
\begin{equation}  \label{4}
\phi _n(y)=-\phi _n(-y)+0\text{mod}\left( 2\pi \right) .
\end{equation}
Note that due to the symmetry relations (\ref{4}) conditions (\ref{3}) alone
do not specify any boundary value problem for (\ref{1}), (\ref{2}): The
additive constants on the right-hand side of (\ref{4}) remain undetermined.
The determination of the constants $0$mod$\left( 2\pi \right) $ that define
a topological type of the solution requires imposition of boundary
conditions on $\phi _n$. A misunderstanding of this issue misled the author
of the Comment into thinking that the solution to the boundary value problem
for static SG equations could be ''nonunique''. However, the actual
mathematical arbitrariness of (\ref{1}), (\ref{2}), related to the
unphysical degree of freedom, lies in an infinite number of these equations:
For an infinite number of ordinary differential equations one cannot
formulate any existence and uniqueness theorems.\cite{HP80} The exact
minimization procedure of Refs. \cite{K99,K01} (or, alternatively, of
Appendix A) leads to relations 
\begin{equation}  \label{5}
\phi _n(y)=\phi _{n+1}(y)\equiv \phi (y)
\end{equation}
that reduce (\ref{1}), (\ref{2}) to the well-defined single static SG
equation, with the single Josephson length $\lambda _J$.\cite{r4}
Unfortunately, in spite of ''extensive theoretical studies''\cite{Kr02} of
this equation, the fact that it admits an exact analytical solution, valid
for any $H$ and $L$, has not been realized in any previous publications: 
\begin{equation}  \label{6}
\phi (y)=\pi \left( N_v-1\right) +2\text{am}\left( \frac y{k\lambda _J}%
+K\left( k^2\right) ,k\right) ,
\end{equation}
\begin{equation}  \label{7}
\text{dn}\left( \frac L{k\lambda _J},k\right) =\frac{\sqrt{1-k^2}}k\frac{H_s}%
H,\quad N_v=2m,\quad m=0,1,\ldots ;
\end{equation}
\begin{equation}  \label{8}
\phi (y)=\pi N_v+2\text{am}\left( \frac y{k\lambda _J},k\right) ,
\end{equation}
\begin{equation}  \label{9}
\text{dn}\left( \frac L{k\lambda _J},k\right) =k\frac H{H_s},\quad
N_v=2m+1,\quad m=0,1,\ldots ,
\end{equation}
where am$\left( u\right) $ and dn$\left( u\right) =\frac d{du}$am$\left(
u\right) $ are the Jakobi elliptic functions, and $K\left( k^2\right) $ is
the elliptic integral of the first kind. The superheating (penetration)
field $H_s\equiv H_{s\infty }=\left( ep\lambda _J\right) ^{-1}$ determines
the upper bound of the existence of the Meissner state in the semiinfinite
(along the layers) system $0\leq y<+\infty $.\cite{K99,K01} The topological
number\cite{R82} $N_v=0,1,\ldots $ specifies the number of vortex planes,
with $N_v=0$ for the Meissner state. The range of the existence of the
soliton solution with $N_v\geq 1$ is given by 
\begin{equation}  \label{10}
\sqrt{H_{N_v-1}^2-H_s^2}\leq H\leq H_{N_v},
\end{equation}
where $H_{N_v}$ is determined by the implicit equation 
\begin{equation}  \label{11}
\frac L{\lambda _J}=\left( N_v+1\right) \frac{H_s}{H_{N_v}}K\left( \frac{%
H_s^2}{H_{N_v}^2}\right) ,
\end{equation}
with $H_0\equiv H_{sL}>H_s$ being the superheating (penetration) field for $%
L<\infty $. The lower bound in (\ref{10}) (the appearance of the soliton) is
determined by applying (\ref{3}) to the solution of the boundary value
problem with topological conditions 
\begin{equation}  \label{12}
\phi (-L)=0,\quad \phi (0)=\pi N_v,
\end{equation}
which corresponds to the requirement that the density of the Josephson
energy [and, essentially, the free energy: see (\ref{a4})] be a minimum at
the boundaries $y=\pm L$. [Owing to the symmetry relations (\ref{4}), it is
convenient to impose a topological condition at $y=0$ instead of $y=+L$. The
role of topological boundary conditions is exhaustively discussed in the
literature on soliton physics.\cite{R82}] The topological solution becomes
unstable, when the density of the Josephson energy is a maximum at $y=\pm L$%
, i.e., 
\begin{equation}  \label{13}
\phi (-L)=-\pi ,\quad \phi (0)=\pi N_v,
\end{equation}
which determines, by (\ref{3}), the upper bound in (\ref{10}). [For the
Meissner solution, an analogous condition, determining $H_{sL}$, is obtained
from (\ref{13}) by setting $N_v=0$.] In the intermediate field range, the
appropriate boundary conditions follow from (\ref{3}) and continuous
dependence on $H$: 
\begin{equation}  \label{14}
\frac{d\phi }{dy}\left( -L\right) =2epH,\quad \phi (0)=\pi N_v.
\end{equation}
[For the Meissner solution, we should set $N_v=0$ in (\ref{14}).]

Under corresponding redefinition of the parameters $\lambda _J$, $H_s$, the
solution (\ref{6})-(\ref{11}) describes the single junction ($N=2$) and the
double-junction stack ($N=3$, see section II), which shows that vortex
planes are a direct generalization of ordinary Josephson vortices.\cite
{OS67,K70} (In the single-junction case, $N_v$ should be interpreted as the
number of ordinary vortices.) Equations (\ref{6})-(\ref{11}) cover all
well-known limiting cases.\cite{K70,K99,K01}

An important feature of the exact solution is an overlap of the regions (\ref
{10}) with different topological numbers $N_v$. For the single junction,
this overlap (not the alleged\cite{Kr02} ''nonuniqueness of the solution'')
was first established by a numerical analysis.\cite{OS67} It is also
qualitatively discussed in Ref. \cite{K70} . Mathematically, the overlap is
related to the fact that the solution with $N_v$ cannot be continuously
transformed into the solution with $N_v+1$ by changing $H$. As shown in
section II, the overlap is typical of vortex-plane solutions.

For $N<\infty $, equations (\ref{1}) still hold, however, condition (\ref{2}%
) should be replaced by the conditions 
\begin{equation}  \label{15}
n=1,\ldots ,N-1,\quad \phi _0=\phi _N\equiv 0\quad \left( 2\leq N<\infty
\right) .
\end{equation}
Relations (\ref{3}), (\ref{4}) also hold, however, relations (\ref{5})
should be replace by 
\begin{equation}  \label{16}
\phi _n=\phi _{N-n},
\end{equation}
obtained from the boundary conditions on the local field at $x=0,$ $x=\left(
N-1\right) p$.\cite{K0,K2} In contrast to the case $N=\infty $, we can now
take advantage of powerful analytical methods of the theory of ordinary
differential equations. Using Picard's uniqueness and existence theorem,\cite
{HP80} we establish the central analytical property of (\ref{1}), (\ref{15}):

{\bf Lemma}. The initial value problem for (\ref{1}), (\ref{15}) with
arbitrary initial conditions $\phi _n(y_0)=\alpha _n$, $\frac{d\phi _n}{dy}%
(y_0)=\beta _n$ has a unique solution in the whole interval $-\infty
<y<+\infty $. This solution has continuous derivatives with respect to $y$
of arbitrary order and continuously depends on the initial data.

This fundamental existence and uniqueness Lemma will allow us to obtain a
complete classification of all soliton (vortex) solutions to (\ref{1}), (\ref
{15}).

\section{The classification of soliton (vortex) solutions to static coupled
SG equations for $N<\infty $}

By analogy with (\ref{6})-(\ref{11}), we should find solutions to (\ref{1}),
(\ref{15}) subject to topological boundary conditions 
\begin{equation}  \label{17}
\phi _n(-L)=0,\quad \phi _n(0)=0\text{mod}\left( \pi \right) ,
\end{equation}
with all possible nonzero sets of the constants $0$mod$\left( \pi \right) $,
compatible with the requirement (\ref{3}) that all $\frac{d\phi _n}{dy}%
\left( -L\right) $ be equal to each other. This task can be reformulated in
terms of the initial value problem at $y=-L$: 
\begin{equation}  \label{18}
\phi _n(-L)=0,\quad \frac{d\phi _n}{dy}\left( -L\right) =2ep\tilde H>0,
\end{equation}
where $\tilde H$ is so far unknown. According to the Lemma of section I, the
solution to (\ref{18}), satisfying topological conditions at $y=0$, will
represent in $\left[ -L,L\right] $ the unique type of the solution to (\ref
{17}), compatible with the boundary conditions on $\frac{d\phi _n}{dy}$.
Using continuous dependence of this solution on $H$, we can establish a full
range of its existence. This program is carried out to a full extent in Ref. 
\cite{K2} . Here we present the main results. As could be expected, the only
possible soliton solutions for $H>0$, $L<\infty $ turn out to be of the
vortex-plane type. In addition to the symmetry relations (\ref{4}), (\ref{16}%
), they satisfy the conditions 
\begin{equation}  \label{20}
-\pi \leq \phi _n\left( -L\right) \leq 0,\quad \phi _n(0)=\pi N_v,\quad 
\frac{d\phi _n}{dy}\left( y\right) >0,\quad n=1,\ldots ,N-1.
\end{equation}
The range of the existence of the solution with $N_v\geq 1$ is given by (\ref
{10}), where the superheating (penetration) field $H_s$ is now an explicit
analytical function of $N$ [$H_s\left( N<\infty \right) >H_s\left( N=\infty
\right) \equiv \left( ep\lambda _J\right) ^{-1}$]$,$\cite{K0,K2} and $%
H_{N_v} $ is determined by the solution to the boundary value problem 
\begin{equation}  \label{20.1}
\phi _n\left( -L\right) =-\pi ,\quad \phi _n(0)=\pi N_v,\quad n=1,\ldots
,N-1.
\end{equation}
Using analogous considerations, it is straightforward to establish the
absence of any soliton solutions (including the vortex planes ) for $H=0$, $%
L<\infty $. (The latter result is indirectly confirmed by the numerical
simulations of Refs. \cite{SBP93,Kr01} for $H=0$, performed in a finite
interval on the $y$ axis: Appropriate topological boundary conditions are
not fulfilled in the figures presented therein.)

As a simple illustration of the above mathematical techniques, we present an
explicit proof of the absence of single-vortex solutions to a pair of
coupled static SG equations ($N=3$) for $H>0$. In this case, the boundary
value problem for a single vortex, positioned at $n=1$, is formulated as
follows: 
\begin{equation}  \label{21}
\phi _{1,2}\left( -L\right) =0,\quad \phi _1\left( 0\right) =\pi ,\quad \phi
_2\left( 0\right) =0.
\end{equation}
However, this boundary value problem has no solution, meeting the
requirement $\frac{d\phi _1}{dy}\left( -L\right) =\frac{d\phi _2}{dy}\left(
-L\right) $, for any $\tilde H>0$. Indeed, the initial value problem (\ref
{18}) for $N=3$ has an exact analytical solution on the whole axis $-\infty
<y<+\infty $, valid for any $\tilde H>0$: 
\begin{equation}  \label{22}
\phi _1(y)=\phi _2(y)\equiv \phi (y)=-\pi +2\text{am}\left( \frac{y+L}{%
k\lambda _{J1}}+K\left( k^2\right) ,k\right) ,
\end{equation}
\[
\lambda _{J1}=\frac{\epsilon \lambda _J}{\sqrt{1+\epsilon ^2}},\quad k=\frac %
1{\sqrt{1+\left( ep\lambda _{J1}\tilde H\right) ^2}}, 
\]
where $\lambda _{J1}$ is the unique Josephson length for $N=3$. (Contrary to
a remark in the Comment, both the Meissner solution and the vortex planes
for $3\leq N<\infty $ are characterized by a distribution of Josephson
lengths, whose number is equal to the integer part of $\frac N2$.\cite{K0,K2}%
) According to the Lemma, expression (\ref{22}) is the unique solution,
compatible with the boundary conditions at $y=-L$ for any $\tilde H>0$. As
anticipated by (\ref{20}), it satisfies the condition $\frac{d\phi }{dy}>0$
on the whole axis $y$, whereas (\ref{3}) and (\ref{21}) imply $\frac{d\phi _2%
}{dy}\left( 0\right) <0$. Note that although the equality $\phi _1=\phi _2$
could be expected by (\ref{16}), these symmetry relations were not employed
in our proof. (The fact that $\phi _1=\phi _2$ for $H>0$ was verified
experimentally on artificial double-junction stacks.\cite{N96})

Although vortex planes are unique soliton solutions for $H>0$, $L<\infty $,
the situation changes drastically at $H=0$, $L=\infty $, when there is no
need to bother about boundary conditions on $\frac{d\phi _{n}}{dy}$. The
imposition of topological boundary conditions 
\begin{equation}
\phi _{n}(y)%
%TCIMACRO{\underset{y\rightarrow -\infty }{\rightarrow }}%
%BeginExpansion
\mathrel{\mathop{\rightarrow }\limits_{y\rightarrow -\infty }}%
%EndExpansion
0,\quad \phi _{n}(y)=0\ \text{mod}(\pi )  \label{24}
\end{equation}
automatically ensures the fulfillment of the asymptotic boundary conditions 
\begin{equation}
\frac{d\phi _{n}}{dy}\left( y\right) 
%TCIMACRO{\underset{y\rightarrow -\infty }{\rightarrow }}%
%BeginExpansion
\mathrel{\mathop{\rightarrow }\limits_{y\rightarrow -\infty }}%
%EndExpansion
0,  \label{25}
\end{equation}
by virtue of Eqs. (\ref{1}), (\ref{15}) themselves and some elementary
theorems of mathematical analysis. As shown in Refs. \cite{K0,K2} , apart
from the vortex-plane solution, equations (\ref{1}), (\ref{15}) admit for $%
H=0$, $L=\infty $ a variety of topological configurations: single-vortex,
vortex-vortex and vortex-antivortex solutions. However, all these solutions
are subject to the requirement that the constants $0$mod$\left( \pi \right) $
be chosen from the set $0,\pm \pi $, i.e., each $\phi _{n}$ can
''accommodate'' for $H=0$, $L=\infty $ no more than one vortex or
antivortex. (This is a typical result of the theory of nonlinear ordinary
differential equations that could not be obtained by means of numerical
simulations,\cite{SBP93,KrW97,Kr01} restricted to a finite interval on the $y
$ axis.) In particular, the solution representing a single vortex,
positioned at $n=l$, is specified by the conditions 
\[
\phi _{l}(y)%
%TCIMACRO{\underset{y\rightarrow -\infty }{\rightarrow }}%
%BeginExpansion
\mathrel{\mathop{\rightarrow }\limits_{y\rightarrow -\infty }}%
%EndExpansion
0,\quad \frac{d\phi _{l}}{dy}\left( y\right) 
%TCIMACRO{\underset{y\rightarrow -\infty }{\rightarrow }}%
%BeginExpansion
\mathrel{\mathop{\rightarrow }\limits_{y\rightarrow -\infty }}%
%EndExpansion
0,\quad \phi _{l}(0)=\pi \text{,}\quad \frac{d\phi _{l}}{dy}\left( 0\right)
>0\text{,}
\]
\begin{equation}
\phi _{n}(y)%
%TCIMACRO{\underset{y\rightarrow -\infty }{\rightarrow }}%
%BeginExpansion
\mathrel{\mathop{\rightarrow }\limits_{y\rightarrow -\infty }}%
%EndExpansion
0,\quad \frac{d\phi _{n}}{dy}\left( y\right) 
%TCIMACRO{\underset{y\rightarrow -\infty }{\rightarrow }}%
%BeginExpansion
\mathrel{\mathop{\rightarrow }\limits_{y\rightarrow -\infty }}%
%EndExpansion
0,\quad \phi _{n}(0)=0\text{,}\quad \frac{d\phi _{n}}{dy}\left( 0\right) <0%
\text{, for }n\neq l,  \label{27}
\end{equation}
plus a certain existence condition on all $\frac{d\phi _{n}}{dy}\left(
0\right) $.\cite{K0,K2} The main{\it \ }distinctive feature of this
solution, the negative sign of $\frac{d\phi _{n\neq l}}{dy}\left( 0\right) $
in (\ref{27}), is clearly reproduced by figures 1, 2 in Ref. \cite{Kr01} .
In this context, the remark\cite{Kr02} that the numerical solution\cite{Kr01}
''agrees well with Bulaevskii\cite{B73} and Clem-Coffey\cite{CC90}
solutions'' is inappropriate: Instead of exact Eqs. (\ref{1}), references 
\cite{B73,CC90} employ a mathematically ill-defined ''continuum-limit
approximation'' that completely neglects this feature. (As shown by Farid,
\cite{F98} ''solutions'' of this type are, in fact, a mathematical fiction.)

Summarizing, the actual domain of the existence of the single-vortex
solution is restricted to $H=0$, $L=\infty $. Concerning figure 1 in the
Comment, based on the numerical calculations\cite{Kr01} for $H=0$, the ratio
of the self-energy of the vortex plane to that of the single vortex at $H=0$
is unimportant, in light of the absence of the latter at $H>0$. For the same
reason, the expression\cite{Kr02} ''$H_{c1}=4e\pi E_{\text{{\it sing}}}/\Phi
_0$'' is physically senseless. The actual lower critical field $H_{c1}$ is
determined by the condition that the Gibbs free energy of the state with a
single vortex plane be equal to that of the Meissner state. In the two
exactly solvable cases, $N=\infty $ and $N=3$ with $(epH_s)^{-1}\ll L<\infty 
$, it is given by $H_{c1}=\frac 2\pi H_s$, as for the single junction.\cite
{K70} (In general, $\sqrt{H_{sL}^2-H_s^2}<H_{c1}<H_{sL}$.) Calculations\cite
{KrW97,Kr00,Kr02} of the number of ''quasiequilibrium fluxon modes'' by
means of combinatorics are also invalid, because they do not take into
account the necessity to satisfy topological boundary conditions together
with the boundary conditions on $\frac{d\phi _n}{dy}$. (Figure 6 in Ref. 
\cite{Kr00} does not provide any evidence that such conditions for isolated
fluxons are fulfilled in the numerical simulations, either.) The comparison
\cite{Kr02} of the soliton vortex-plane solutions with the laminar model\cite
{G64,dG} for type-II superconductors is a misunderstanding: Alternating
superconducting and normal layers, envisaged by this model, have nothing to
do with soliton physics and do not possess the property of topological
stability.

The statement\cite{Kr02} that ''free energy of any isolated solution is
twice the Josephson energy'' is incorrect: See, e.g., (\ref{a4}). However,
the self-energy of soliton solutions for $H=0$, $L=\infty $ is indeed twice
the Josephson energy, which has been established by the use of the first
integral of (\ref{1}) in Ref. \cite{K0} (not Ref. \cite{KrW97} , as is
claimed in the Comment). (From a field-theoretical point of view, this fact
is a manifestation of the virial theorem.\cite{R82})

\section{Josephson-vortex penetration}

Isolated Josephson vortices cannot penetrate layered superconductors and
stacked junctions at any $H>0$, because they do not form any equilibrium
configurations.\cite{K99,K01} This problem cannot be circumvented by any
consideration of penetration as a ''dynamic process''\cite{Kr02} whose final
stage should be an equilibrium flux configuration. The experiments on
artificial stacks\cite{N96,Yu98} have clearly shown that Josephson vortices
penetrate all the junctions simultaneously{\it \ }and coherently (i.e., in
the form of vortex planes), in complete agreement with the scenario of Refs. 
\cite{K99,K01} .

A description in terms of static equations, accepted in Refs. \cite{K99,K01}
and previous publications\cite{OS67,K70} on single Josephson junctions,
implies that a flux relaxation time is much smaller than a characteristic
time of the change of the external field, which corresponds to experimental
conditions of thermodynamic equilibrium. As in the case of Abrikosov
vortices in type-II superconductors,\cite{dG} penetration occurs when vortex
planes appear at the surface $y=\pm L$ (i.e., at $H=H_{N_v}$), which
signifies the vanishing of the surface barrier. (In contrast, the static SG
equations with $N\geq 3$ do not admit solutions with a single Josephson
vortex at the surface.) Mathematically, penetration constitutes a change of
the topological type of the solution and should be interpreted as a series
of first-order phase transition.\cite{K70,K99,K01} Discontinuities related
to these phase transitions manifest themselves, in particular, in jumps of
magnetization.\cite{K70,K99,K01} Such jumps of magnetization were indeed
observed experimentally in stacked junctions.\cite{Yu98}

What figure 2 in the Comment actually shows is a violation of the conditions
of thermodynamic equilibrium in the numerical simulations. The hypothesis
\cite{Kr02} of 'decomposition of a ''breather''' is also irrelevant.
Certainly, the single ($N=2$) time-dependent SG equation for $H=0$, $%
L=\infty $ has an exact analytical solution, known as the ''breather''.\cite
{R82} However, apart from the fact that time-dependent SG equations have
nothing to do with thermodynamic equilibrium, the author of the Comment does
not demonstrate the existence of such a solution for $N\geq 3$, $H>0$, $%
L<\infty $ and does not explain why it should ''decompose''.

\section{Experimental observations of Josephson-vortex configurations: Some
concluding remarks}

There is no contradiction between our conclusion about the absence of
single-vortex configurations at $H>0$ and the observations\cite{Mo98} of
non-equilibrium isolated vortices in layered high-$T_c$ superconductors at $%
H=0$. In the case of Ref. \cite{Mo98} , isolated vortices can be pinned by
structural defects, because their energy is lower and the $c$-axis extent is
smaller than those of vortex planes at $H=0$.\cite{K01}

Measurements\cite{Kr00} of $c$-axis transport properties do not provide any
direct information on flux distribution in stacked junctions. Both the
multivaluedness and the aperiodicity of $I_c(H)$ dependence, reported in
Ref. \cite{Kr00} , can be explained by the overlap of states with different
numbers of vortex planes, discussed in section I.

The Josephson-vortex structure in artificial stacked junctions for $H>0$, $%
I=0$ was directly observed in Ref. \cite{N96} (by low-temperature scanning
electron microscopy) and in Ref. \cite{Yu98} (by polarized neutron
reflection). As is emphasized in the introduction and sections II, III, all
our theoretical conclusions fully agree with these observations. In
contrast, the concept of ''isolated fluxons''\cite{Kr02} stands in direct
contradiction to the experiments of Refs. \cite{N96,Yu98} .

In conclusion, a revision of some old prejudices in the theory of Josephson
systems should begin with the case of the single junction ($N=2$). In spite
of ''extensive theoretical studies''\cite{Kr02} for almost four decades,
there is still an erroneous belief\cite{BCG91,Kr01} that Josephson vortices
''do not exist'' in single junctions with $L\ll \lambda _J$, which is
refuted by the exact, closed-form solution (\ref{6})-(\ref{11}). [As shown
in Refs. \cite{K99,K01} , exactly the ordinary Josephson vortices and the
vortex planes account for the occurrence of the well-known Fraunhofer
pattern of $I_c(H)$ for small $L.$] The lack of the understanding of
analytical properties of the single SG equation is one of the reasons why
the solution to the more complex coupled SG equations was not obtained in
previous publications. Adequate mathematical techniques to tackle this
problem are proposed in Refs. \cite{K99,K01,K0,K2} . The author of the
Comment himself could profit from the application of these techniques to his
own research.

\appendix

\section{A new method of exact minimization of the Gibbs free energy of the
infinite LD model}

Equations (\ref{1}), (\ref{2}) have the first integral 
\begin{equation}  \label{a1}
C\left( H\right) -\sum_n\cos \phi _n(y)=\frac{\epsilon ^2\lambda _J^2}2%
\sum_nG_\infty \left( n,m\right) \frac{d\phi _n(y)}{dy}\frac{d\phi _n(y)}{dy}%
,
\end{equation}
\[
C\left( H\right) =2\left( ep\lambda _JH\right) ^2\frac V{pL}+\sum_n\cos \phi
_n(L), 
\]
\[
G_\infty \left( n,m\right) =\frac{\mu ^{\left| n-m\right| }}{2\epsilon \sqrt{%
1+\frac{\epsilon ^2}4}},\quad \mu =1+\frac{\epsilon ^2}2-\epsilon \sqrt{1+%
\frac{\epsilon ^2}4}. 
\]
The term on the right-hand side of (\ref{a1}) is the density of
electromagnetic energy. For $r(T)\ll 1$ and $H\ll H_{c2}$, one can eliminate
the electromagnetic-energy term from the LD free-energy functional\cite{K01}
by the use of the Maxwell equations and (\ref{a1}), obtaining
\[
\Omega _{LD}\left[ \varphi _n;H\right] =\frac{H_c^2(T)V}{4\pi }\left[ -\frac %
12+r(T)\left[ 1+4\left( ep\lambda _JH\right) ^2+\frac{2pL}V\sum_n\cos \phi
_n\left( L\right) \right. \right. 
\]
\begin{equation}  \label{a4}
\left. \left. -\frac{2p}V\sum_n\int\limits_{-L}^Ldy\cos \phi _n(y)-\frac{%
4ep^2\lambda _JLH}V\sum_n\left[ \phi _n(L)-\phi _n(-L)\right] \right] \right]
.
\end{equation}
Taking the variation of (\ref{a4}) with respect to $\varphi _n(y)$, one
immediately arrives at the relations (\ref{5}). (Mathematically, the
elimination of the electromagnetic-energy term is equivalent to the
elimination of the intrinsic constraint in Ref. \cite{K01}.) Unfortunately,
the existence of the first integral (\ref{a1}) was not noticed in any
previous publications on the infinite LD model,\cite
{B73,CC90,BC91,BCG91,BLK92,Ko93} which partly explains the difficulties with
the minimization of the Gibbs free energy with respect to $\varphi _n$.


\begin{references}
\bibitem{r4}  Note that Eqs. (\ref{1}), (\ref{2}) for $H=0$, $L=\infty $
possess an exact topological solution that cannot be obtained by means of
the minimization of the Gibbs free energy, namely, an infinite
vortex-antivortex configuration: $\phi _{n}(y)=(-1)^{n}\phi (y)$, $\phi
(y)=4\arctan \exp \left[ \frac{y}{\lambda _{\min }}\right] $, $\lambda
_{\min }=\frac{\epsilon \lambda _{J}}{\sqrt{4+\epsilon ^{2}}}$. However,
this solution vanishes for any $H>0$, $L<\infty $.
\end{references}
\end{document}